# Cryptompress: A Symmetric Cryptography algorithm to deny Bruteforce Attack

Vivek Kumar[1] and Sandeep Sharma[2]
[1,2]Department of Electronics and Communication Engineering,
Dehradun Institute of Technology, Uttarakhand, India, 248009
[1]er_vivek@outlook.com, [2]tek.learn@gmail.com

**Abstract**

Cryptompress, a new 128-bit (initial) private-key cryptography algorithm is proposed. It uses a block size of at least 30 bits and increments prior key size to additional 32 bits on each unsuccessful attempt of any means, including bruteforcing, further changing a specific portion of the cyphertext using the reformed Feistel network. Encryption process results from a proposed compression sequence developed using lookup table and shift operations followed by key generation. Eventually, four matrixes named add-sub matrix, reduced matrix, sequence matrix and term matrix are obtained which ultimately forms a cyphertext.

**Index Terms**

Cryptompress, Symmetric cryptography, Compression, Bruteforcing, Feistel network, Data security, Information security

## I. INTRODUCTION

Cryptography provides a way to transform sensitive information (plaintext) into a secure one (cyphertext) while assuring both data and information security. The term cryptography is originated from two words, "crypto" means "hidden" and "graphy" means "to write". [1] The process gets completed in two steps, encryption and decryption. During encryption, sensitive data is allowed to pass through a series of pre-defined steps involving distinct operations to achieve a cyphertext. However, there's always a need of a key which may be either public or private depending upon the requirement of a system. Relying on the type of key used, cryptography can be divided into two domains: Symmetric Cryptography (SC) and Asymmetric Cryptography (AC). In symmetric cryptography, a secret key is used for both encryption of plaintext and decryption of cyphertext which is shared at both ends. On the flip side, in asymmetric cryptography a public key is used for encryption of plaintext by everyone and a private key for decryption. [2][3]

Various SC algorithms have been designed so far, with the most famous ones being AES [4], RC2 [5], IDEA [6], Blowfish [7] etc. Such major algorithms do use key length of at least 112 bits (exception being DES [8] and some variable key length algorithms). Yet, it is regarded that for a key length smaller or equal to 112 bits, key spaces can be computed using brute force attack. But, once a key reaches above this threshold brute force attack seems irrelevant, as it retains no practical significance. However, with rapid growth in chip technology (big thanks to Moore's Law), bruteforcing such high key space may soon become reality. Hence, an adaptive encryption algorithm needs to be developed to become future ready and retain encryption without concerning about brute force attack.

In this paper, our purpose is to develop a flexible encryption algorithm using 128 bit private key (initially) which currently remains undefiled from brute force attack and increase the key length on each unsuccessful attempt as well as increasing the complexity of the cyphertext. Although there are some complex steps involved during initialization, the overall output is more resilient and robust to any known vulnerabilities.

The paper is structured as follows: we premise our approach in section II in conjunction with the design flow diagram and the working process. Furthermore, review of the compression and decompression technique along with the structure for all four derived matrixes namely Add-Sub Matrix (ASM), Reduced Matrix (RM), Sequence Matrix (SM) and Term Matrix (TM) are also presented in the same. Additionally, this section also covers the description of key generation and structure. Section III describes increasing size nature of the key, along with the reformed Feistel network and its effect on the key structure and cyphertext. We provide a numerical illustration for better clarification in section IV. The paper ends by depicting conclusion and references in section V and VI respectively.

## II. CRYPTOMPRESS: PRINCIPLE OPERATION AND WORKING

### A. Principle operation

Cryptompress is a 128 bit (initial) symmetric key cryptography method. It uses 30 bits of block size to obtain corresponding cyphertext and variable key. It is formulated in three major segments; compression framework, key generation and key expansion accompanied with transforming Sequence Matrix (SM) on each unsuccessful attempt. The flowchart in figure 1 and figure 2 gives an overview regarding the complete layout of the Cryptompress.

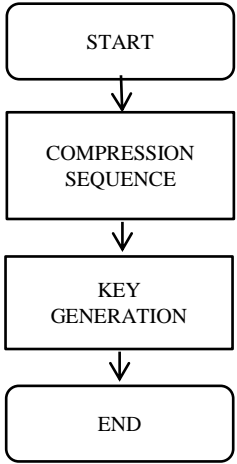

Fig. 1. Encryption layout

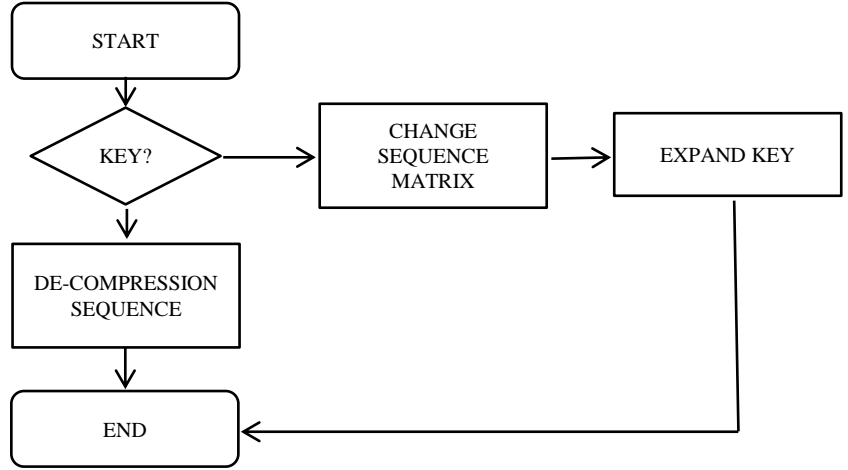

Fig. 2. Decryption layout

B. *The compression framework*

To start with encryption process, we follow a compression methodology which is mostly applicable for data comprising of consecutive similar binary pairs. The compression methodology has been presented along with the encryption phase. We consider 30 bits block code and lay down a mapping function for two consecutive bits i.e. 00, 01, 10 and 11 mapped to first four positive integer prime numbers i.e. 2, 3, 5 and 7. One can express such mapping function M as:

$$M = \begin{cases} 00 \Leftrightarrow 2 \\ 01 \Leftrightarrow 3 \\ 10 \Leftrightarrow 5 \\ 11 \Leftrightarrow 7 \end{cases} \quad (1)$$

This will transform the block code of 30 bits into 15 integers (15 bytes). Next off, the *compression sequence* initiates deploying three main operations, right shift operation, lookup table corresponding to *Add-Sub Matrix (ASM)* and simpler addition/subtraction. Compression sequence has a rigid boundary in a sense that it will begin its procedure by choosing the first occurring prime integer (target integer) of the block. Once, first occurring prime integer is done with, the next occurring integer becomes target integer and so on up till the last integer available in the block code is finished.

Once a target integer is acquired, next operation is to perform right shifting the target integer and follow two rules. First, add similar integer(s) together and feed the sequence serial number ($S_n$) along with the number of redundant integers ($R_n$) summed up into a matrix, known as *Sequence Matrix (SM)*. Second, whenever a different integer is crossed, perform a certain addition/subtraction operation in the target number by looking in the ASM. Execution takes place until the target number reaches the end of the block, finally ending up as target number outcome. This outcome is stored in a new matrix called *Reduced Matrix (RM)*. Similar operations are performed for the rest of the available target prime integers resulting in a transformed data only available in RM, SM and new matrix called *Term Matrix (TM)* which holds the information of the last sequence serial number corresponding to each target number.

Overall, compression sequence results into three matrixes whose total size may be bigger than the original block code depending upon the occurrence of the consecutive similar binary pairs. For instance, if similar binary pairs occur consecutively for a bigger block size then the resultant sequence matrix will contain less information, hence compressing the data effectively.

To obtain data back in its original form (decompression), first off, refer to the last occurring target integer highest sequence number from the TM and assign its respective outcome from RM. Proceed with left shifting using the ASM (which in this case replaces add/sub with sub/add operations) and fetching number of known redundant integers from the SM.

C. *Description of matrixes*

As aforementioned, there are four main matrixes which need to be dealt with during the Cryptompress. Let us describe them here with their structures and significance.

1. **Add-Sum Matrix (ASM):** An Add-Sum Matrix (ASM) works like a *lookup table* during both compression and decompression sequences. It has a major impact on the Reduced Matrix (RM) data as it works like a *channel* between block code data and Reduced Matrix data. It is so relevant because it depicts the type of operation to be performed on a target number when it right/left shifts any distinct integer. It solely determines whether there would be 1 bit addition or 1 bit subtraction to the target number. Table 1 elucidates the structure of ASM, here *Order* depicts the range of values from 0000-1111 which means that for every "0" there would occur 1 bit subtraction (-1) from the target number and for every

"1" there would occur 1 bit addition (+1) to the target number. The process inverts during left shifts. Here, "**X**" denotes addition of the target number by itself and has no relevance to ASM. Refer section IV for illustration learning.

TABLE 1: STRUCTURE OF ASM

| Order | Target Number | 2 (Vert.) | 3 (Vert.) | 5 (Vert.) | 7 (Vert.) |
|---|---|---|---|---|---|
| 0000-1111 | 2 (Hort.) | X | +1/-1 | +1/-1 | +1/-1 |
| 0000-1111 | 3 (Hort.) | +1/-1 | X | +1/-1 | +1/-1 |
| 0000-1111 | 5 (Hort.) | +1/-1 | +1/-1 | X | +1/-1 |
| 0000-1111 | 7 (Hort.) | +1/-1 | +1/-1 | +1/-1 | X |

2. **Sequence Matrix (SM):** During shifting, when a target number finds a similar redundant integer then it sums up all those integers and maintains a Sequence Matrix (SM). SM stores such upcoming sequence serial number along with the number of redundant integers summed up in each sequence. Table 2 depicts the structure of SM. Here, format "$S_n / R_n$" means sequence serial number | number of redundant integers.

TABLE 2: STRUCTURE OF SM

| 2 | 3 | 5 | 7 |
|---|---|---|---|
| $S_1 \mid R_1$ | $S_1 \mid R_1$ | $S_1 \mid R_1$ | $S_1 \mid R_1$ |
| . | . | . | . |
| . | . | . | . |
| $S_n \mid R_n$ | $S_n \mid R_n$ | $S_n \mid R_n$ | $S_n \mid R_n$ |

3. **Reduced Matrix (RM):** Each time a target number completes its block course, it ends up as an *outcome*. All four such outcomes are stored in a matrix, coined as Reduced Matrix (RM). It has a very simple architecture as it consists of only four outcomes corresponding to each target numbers as depicted in Table 3.

TABLE 3: STRUCTURE OF RM

| 2 | 3 | 5 | 7 |
|---|---|---|---|
| $Outcome_2$ | $Outcome_3$ | $Outcome_5$ | $Outcome_7$ |

4. **Term Matrix (TM):** Term Matrix stores the last occurring sequence serial number corresponding to each target number as shown below. Its representation format is $T_x \mid S_n$ where $T_x$ ($x \in 2, 3\ 5, 7$) governs the last occurring prime number in the block.

TABLE 4: STRUCTURE OF TM

| 2 | 3 | 5 | 7 |
|---|---|---|---|
| $T_x \mid S_n$ | $T_x \mid S_n$ | $T_x \mid S_n$ | $T_x \mid S_n$ |

D. *Key generation*

In this section, we deliver the 128 bit private key generation mechanism. So far, we are able to generate a transformed block code data, now let us know how does it is relates with encryption.

The Cryptompress key is divided into 5 sub categories, of which fifth category has a dynamic size and is not available initially. The structure of the key is depicted in table 5.

TABLE 5: STRUCTURE OF KEY

| Keys: | ASM key | RM key | TM key | SM key | STICKY key |
|---|---|---|---|---|---|
| Size: | 48 bits | 16 bits | 16 bits | 48 bits | Dynamic |

Description of the keys:

1. **ASM Key:** With the virtue of ASM key, the complete Add-Sub Matrix (ASM) can be reformed. Using initial 16 bits it defines the type of *order* to be used during the compression-decompression mechanism. Next 16 bits decides the sequential arrangements of the *horizontal target numbers* assigned in ASM according to the mapping function N. Similarly, last 16 bits decides the *vertical target numbers* sequence according to N.

The *mapping function N* distributes all possible combinations of binary values ($2^4$) to each target numbers for their final arrangements in the cyphertext. It works in a cyclic process and transforms the arrangement of target numbers for each matrix as follows:

$$\text{Horizontal ASM} \rightarrow \text{Vertical ASM} \rightarrow \text{RM} \rightarrow \text{SM} \rightarrow \text{TM} \rightarrow \text{back to Hort. ASM}$$

The target numbers swaps their arrangement with the next matrix at a position equal to the decimal equivalent of defined binary combination in the key.

2. **RM key:** The 16 bits of the RM key define the new position of each target numbers arrangements by abiding the cyclic rule of mapping function N.

3. **SM key:** First 16 bits of SM key works same as RM key for Sequence Matrix. The next 32 bits are sub divided into eight sub keys each containing 4 bits which are used to perform Exclusive OR (XOR) function with each factor of target numbers data (using Feistel network). For instance, set of first 4 bits is allowed to do XOR with the left content of prime integer 2 and so on.

4. **TM key:** TM key also has a same feature as RM key.
   **Remark** Term Matrix obtains data in random order because of its data dependency on input. So, it is recommended to map its content to temporary target numbers in sequential state and use them for further mapping.

5. **STICKY key:** STICKY key is the best feature of Cryptompress due to its dynamic nature and reformed Feistel network. This key is not present in the initial run but gets stick with the rest of the key only when an unsuccessful attempt is recognized. On each unsuccessful attempt, it adds 32 more bits to the prior key increasing more complexity to the key. Further explanation to this key will be described in Key expansion section.

Ultimately, the ciphered text will take the format of six columns with rigid sequence given as Order, ASM (hort.), ASM (vert.), RM, SM and TM.

### III. Key Expansion and Feistel Network

So far we have presented the encryption and decryption method. Let us now present about the key expansion and involved reformed Feistel network.

If an attempt to crack initial key becomes unsuccessful then 32 random bits are chosen which acts like a sticky key and further gets attached with the initial key. Next off, these 32 bits are divided into 8 equal sets of 4 bits each. These 8 sets are used as one of the inputs for the Feistel network to be implemented on Sequence Matrix (S.M.).

A Feistel network [7] is a type of cryptography method which divides plaintext into two equal halves (namely left and right), apply a function F including a sub-key on left halve using XOR operation and then swapping the two halves. This process can be repeated into any number of iterations. We, however in this paper presents a reformed Feistel network whose sub-keys are 8 sets of sticky key.

Since, a SM constitutes data in the format $S_n | R_n$ corresponding to each four prime integers. Therefore, first and second set of the sticky key is XOR with each $S_n | R_n$ of the prime number "2" respectively and later their positions are swapped ($\leftrightarrow$) as shown below.

For prime number 2,

$$S_1 \text{ XOR Sub key 1} \quad \leftrightarrow \quad R_1 \text{ XOR Sub key 2}$$
$$\vdots \quad \quad \quad \quad \vdots$$
$$S_n \text{ XOR Sub key 1} \quad \leftrightarrow \quad R_n \text{ XOR Sub key 2}$$

Similarly, sub key 3-8 are XOR on the data of next prime numbers, later performing swapping to obtain a new set of cyphertext.

### IV. An Illustration

The following illustration is meant to give an in depth clarification on the working of Cryptompress.

Before preceding the compression sequence, first step is to form ASM. This will be generated automatically, for instance consider table 6 as ASM.

TABLE 6: ASM

| Order | Target Number | 2 | 3 | 5 | 7 |
|---|---|---|---|---|---|
| 0010 | **2** | X | -1 | +1 | -1 |
| 0011 | **3** | -1 | X | +1 | +1 |
| 0101 | **5** | -1 | +1 | **X** | +1 |
| 0111 | **7** | -1 | +1 | +1 | **X** |

Next off, consider a block code of 30 bits and map the prime numbers according to equation 1.

| 10 | 10 | 10 | 11 | 11 | 01 | 10 | 00 | 11 | 10 | 00 | 11 | 11 | 10 | 01 |
|---|---|---|---|---|---|---|---|---|---|---|---|---|---|---|
| **5** | 5 | 5 | 7 | 7 | 3 | 7 | 2 | 7 | 5 | 2 | 7 | 7 | 5 | 3 |

Here, target integer is 5, so initiating with compression sequence for 5 followed with 7, 3 and later 2. This is shown in table 7, while going through this process keep updating SM and TM respectively.

TABLE 7: COMPRESSION SEQUENCE

| | 5 | 5 | 5 | 7 | 7 | 3 | 7 | 2 | 7 | 5 | 2 | 7 | 7 | 5 | 3 |
|---|---|---|---|---|---|---|---|---|---|---|---|---|---|---|---|
| 5.1 | **15** | | | 7 | 7 | 3 | 7 | 2 | 7 | 5 | 2 | 7 | 7 | 5 | 3 |
| 5.2 | 7 | **16** | 7 | 3 | 7 | 2 | 7 | 5 | 2 | 7 | 7 | 5 | 3 | | |
| 5.3 | 7 | 7 | **17** | 3 | 7 | 2 | 7 | 5 | 2 | 7 | 7 | 5 | 3 | | |
| 5.4 | 7 | 7 | 3 | **18** | 7 | 2 | 7 | 5 | 2 | 7 | 7 | 5 | 3 | | |
| 5.5 | 7 | 7 | 3 | 7 | **19** | 2 | 7 | 5 | 2 | 7 | 7 | 5 | 3 | | |
| 5.6 | 7 | 7 | 3 | 7 | 2 | **18** | 7 | 5 | 2 | 7 | 7 | 5 | 3 | | |
| 5.7 | 7 | 7 | 3 | 7 | 2 | 7 | **19** | 5 | 2 | 7 | 7 | 5 | 3 | | |
| 5.8 | 7 | 7 | 3 | 7 | 2 | 7 | **24** | | 2 | 7 | 7 | 5 | 3 | | |
| 5.9 | 7 | 7 | 3 | 7 | 2 | 7 | 2 | **23** | 7 | 7 | 5 | 3 | | | |
| 5.10 | 7 | 7 | 3 | 7 | 2 | 7 | 2 | 7 | **24** | 7 | 5 | 3 | | | |
| 5.11 | 7 | 7 | 3 | 7 | 2 | 7 | 2 | 7 | 7 | **25** | 5 | 3 | | | |
| 5.12 | 7 | 7 | 3 | 7 | 2 | 7 | 2 | 7 | 7 | **30** | | 3 | | | |
| 5.13 | 7 | 7 | 3 | 7 | 2 | 7 | 2 | 7 | 7 | 3 | **31** | | ←Saved in RM | | |
| 7.1 | **14** | | 3 | 7 | 2 | 7 | 2 | 7 | 7 | 3 | | | | | |
| 7.2 | 3 | **15** | 7 | 2 | 7 | 2 | 7 | 7 | 3 | | | | | | |
| 7.3 | 3 | **22** | | 2 | 7 | 2 | 7 | 7 | 3 | | | | | | |
| 7.4 | 3 | 2 | **21** | 7 | 2 | 7 | 7 | 3 | | | | | | | |
| 7.5 | 3 | 2 | **28** | | 2 | 7 | 7 | 3 | | | | | | | |
| 7.6 | 3 | 2 | 2 | **27** | 7 | 7 | 3 | | | | | | | | |
| 7.7 | 3 | 2 | 2 | **41** | | 7 | 3 | | | | | | | | |
| 7.8 | 3 | 2 | 2 | 3 | **42** | | | ←Saved in RM | | | | | | | |
| 3.1 | 2 | **2** | 2 | 3 | | | | | | | | | | | |
| 3.2 | 2 | 2 | **1** | 3 | | | | | | | | | | | |
| 3.3 | 2 | 2 | **4** | | | | ←Saved in RM | | | | | | | | |
| 2.1 | **4** | | | | | ←Saved in RM | | | | | | | | | |

TABLE 8: SM, RM AND TM

| Target Numbers | SM | RM | TM |
|---|---|---|---|
| 2 | 1\|1 | 4 | 2\|1 |
| 3 | 3\|1 | 4 | 3\|3 |
| 5 | 1\|2 ; 8\|1 ; 12\|1 | 31 | 7\|8 |
| 7 | 1\|1 ; 3\|1 ; 5\|1 ; 7\|2 | 42 | 5\|13 |

**Key formation:** Once, compression sequence is finished, next step is to form a key and transform above resulted data into an encrypted one. The key selection is governed randomly; therefore let us assume a 128 bit key to achieve encryption.

{[0x2357] [0x324A] [0x9153]} {0xDCB5} {0x5124} {[0x3278] [0x12345678]}

Using above key, encrypted data is shown in table 9:

TABLE 9: ENCRYPTED DATA

| Order | ASM (hort.) | ASM (vert.) | RM | SM | TM |
|---|---|---|---|---|---|
| 0010 | 6\|9 ; 4\|9 ; 10\|9 ; 0\|10 | -1X+1+1 | X-1+1-1 | -1+1+1X | 4 |
| 0011 | 4\|4 ; 13\|7 ; 9\|7 | 42 | 4 | X-1-1-1 | 2\|1 |
| 0101 | -1+1X-1 | 0\|5 | -1+1+1X | 5\|13 | 7\|8 |
| 0111 | -1X+1+1 | +1+1X+1 | 31 | 0\|3 | 5\|13 |

## V. Conclusion

In this paper we have discussed about the Cryptompress and mentioned limitation of existing symmetric cryptography algorithms which can be resolved using brute-force attack in future. We presented that how does a cryptography method can involve compression methodology to become much more resilient and hard to break. Cryptompress shows a robust way to overcome Bruteforce attack using its sticky key and strong encryption model. Currently, we are exploring several possible ways to improve its time complexity, initial bit size reduction and transform its core into a parallel processing paradigm.